\documentclass[aps,pra,twocolumn,superscriptaddress]{revtex4}

\usepackage{amsmath,amsfonts,graphicx,color}
\newcommand{\ex}[1]{\mathrm{e}^{\mathrm{i}#1}}
\newcommand{\ket}[1]{|#1\rangle}
\newcommand{\bra}[1]{\langle #1|}
\newcommand{\braket}[2]{\langle #1|#2 \rangle}

\begin{document}
\title{Mutually unbiased bases for the rotor degree of freedom}
\author{Xin \surname{L\"u}}
\affiliation{Centre for Quantum Technologies, %
National University of Singapore, 3 Science Drive 2, 117543, Singapore}
\affiliation{Department of Physics, %
National University of Singapore, 2 Science Drive 3, 117542, Singapore}
\author{Philippe \surname{Raynal}}
\affiliation{Centre for Quantum Technologies, %
National University of Singapore, 3 Science Drive 2, 117543, Singapore}
\author{Berthold-Georg \surname{Englert}}
\affiliation{Centre for Quantum Technologies, %
National University of Singapore, 3 Science Drive 2, 117543, Singapore}
\affiliation{Department of Physics, %
National University of Singapore, 2 Science Drive 3, 117542, Singapore}

\begin{abstract}
We consider the existence of a continuous set of mutually unbiased bases for
the continuous and periodic degree of freedom that describes motion on a
circle (rotor degree of freedom). 
By a singular mapping of the circle to the line, we find a first, but somewhat
unsatisfactory, continuous set which does not relate to an underlying
Heisenberg pair of complementary observables. 
Then, by a nonsingular mapping of the discrete angular momentum basis of the
rotor onto the Fock basis for linear motion, we construct such a Heisenberg
pair for the rotor and use it to obtain a second, fully satisfactory, set of
mutually unbiased bases. 
\end{abstract}

\makeatletter
\renewcommand{\Dated@name}{Posted on the arXiv on }
\makeatother
\date{23 May 2012}

\maketitle

\section{Introduction}\label{sec 1}
Two orthonormal bases of a Hilbert space are called unbiased if the transition
probability from any state of the first basis to any state of the second basis
is independent of the two chosen states. 
In particular, in a $d$-dimensional Hilbert space, two orthonormal bases
$\left\{\ket{a_1},\ket{a_2},\dots,\ket{a_d}\right\}$ and
$\left\{\ket{b_1},\ket{b_2},\dots,\ket{b_d}\right\}$ are
unbiased if 
\begin{equation}
\bigl|\braket{a_i}{b_j}\bigr|^2
=\frac{1}{d}\quad\mbox{for all $i,j=1,2,\dots,d$}.
\end{equation}
A set of Mutually Unbiased Bases (MUB) consists of bases that are pairwise
unbiased.  
In addition to playing a central role in quantum kinematics, MUB provide a
wide range of applications, such as quantum state
tomography~\cite{ivanovic81,wootters89}, quantitative wave-particle duality in
multi-path interferometers~\cite{ekkc08}, quantum key
distribution~\cite{gisin02}, quantum teleportation and dense coding
\cite{durt04b,klimov09,revzen09}.  

For $d$-dimensional spaces, there can be at most $d+1$ MUB, 
and there exist systematic methods for constructing such a
maximal set of MUB in prime-power dimensions (see, for example, Refs.\ 
\cite{ivanovic81,wootters89,bandyopadhyay02,klappenecker04,durt04a,debz10}). 
For other finite dimensions, maximal sets of MUB are unknown. 
Even in the simplest case of dimension six, this is an open problem
although there is quite strong numerical evidence that no more than three MUB
exist~\cite{grassl04, butterley07, brierley08, philippe11}. 
Remarkably, it is always possible to construct a set of three MUB in 
finite-dimensional spaces (see Ref.~\cite{debz10} and references therein).  

More recently, 
the problem of the existence of MUB in the infinite-dimensional case,
that is ${d\to\infty}$, has been addressed. 
This limit is taken by considering a basic Weyl pair \cite{note:pairs} 
of complementary observables whose eigenbases are conjugated 
(Fourier transforms of each other)~\cite{schwinger60}.  
These conjugated eigenbases are
unbiased, and as a manifestation of Bohr's principle of
complementarity~\cite{bohr27}, each Weyl pair is algebraically
complete~\cite{schwinger60} 
as it suffices for a complete parameterization of the degree of freedom. 

For infinite-dimensional spaces, different Weyl pairs corresponding to
different continuous degrees of freedom can be obtained, since there exist
different ways of taking the ${d\to\infty}$
limit~\cite{schwinger01,englert06,debz10}. 
If we treat the Weyl pair symmetrically when taking the limit, then we will
obtain the Weyl pair of complementary observables of the linear motion, that
is, the Heisenberg pair of position observable $Q$ and momentum observable
$P$.

Three different asymmetric ways of taking the  ${d\to\infty}$ limit
produce the basic continuous degrees of freedom of other kinds:
the degrees of freedom (i) of the rotor (described by the
$2\pi$-periodic angular position, and the angular momentum which takes on all
integer values), (ii) of the radial motion (position limited to positive
values, and the momentum takes on all real values), and (iii) of the motion
within a segment (position limited to a finite range, but without periodicity,
and the momentum takes on all real values). 
The corresponding limit ${d\to\infty}$ of a complete set of MUB for
prime dimensions yields a continuous set of MUB for any continuous degree
of freedom \emph{except} for the rotor. Furthermore, these continuous sets of
MUB are related to an underlying Heisenberg pair of complementary
observables. 
This matter is reviewed in sections 1.1.7--1.1.11 of Ref.~\cite{debz10}.

In fact, all the standard methods of constructing a complete set of MUB fail
for the rotor.  
For example, the technique of expressing the MUB as quadratic complex Gaussian
wave functions does not generate more than two MUB. 
Moreover, it is impossible to supplement the two unbiased
bases of the Weyl pair of the rotor with a third unbiased basis. 
The rotor is a very peculiar degree of freedom: It is the only
case where the existence of three MUB has remained unclear.  

The question of the existence of more than two MUB for the rotor was raised in
Ref.~\cite{debz10}, and the aim of this paper is to give an affirmative answer
by constructing a satisfactory continuous set of MUB. 
Indeed, by a rather simple procedure, a first continuous set can be
constructed. 
However, this set is not fully satisfactory since it cannot be related to an
underlying Heisenberg pair of complementary observables as it is the case for
the three other continuous degrees of freedom. 
To get around this discrepancy, we construct a Heisenberg pair of
complementary observables and use it to obtain a second and more suitable
continuous set of MUB. 
This shows that the rotor degree of freedom really is on equal footing with
all the other continuous degrees of freedom. 
The two sets of MUB are found by mapping --- in two different ways --- the
rotor problem onto the well-studied case of linear motion so that the known
method of constructing a continuous set of mutually unbiased bases can then be
applied.

Here is a brief outline of the paper.
In Sec.~\ref{sec 2}, we describe the rotor degree of freedom and repeat the
argument of Ref.~\cite{debz10} that shows explicitly that the two bases
corresponding to the Weyl pair cannot be supplemented with a third unbiased
basis.  
In Sec.~\ref{sec 3}, we provide a first but unsatisfactory continuous set of
MUB for the rotor degree of freedom, with technical details presented in the
Appendix. 
Therefore, in Sec.~\ref{sec 4}, we construct a Heisenberg pair of
complementary observables for the rotor, and use it to find a suitable
continuous set of MUB in Sec.~\ref{sec 5}. 
We close with a summary.

\section{The rotor degree of freedom}\label{sec 2}
A quantum rotor is parameterized by the $2\pi$-periodic angular position and
the angular momentum. 
We denote the hermitian angular-momentum operator by $L$, 
its integer eigenvalues by
$l$, and the corresponding eigenkets and eigenbras by $\ket{l}$ and $\bra{l}$, 
such that~\cite{note:conventions}
\begin{equation}
L\ket{l}=\ket{l}l\quad\text{for\ } l=0,\pm1,\pm2,\dots
\end{equation}
with the orthogonality and completeness relations
\begin{equation}\label{ang mom}
\braket{l}{l'}=\delta_{l,l'}\quad\text{and}\quad 
\sum_{l=-\infty}^{\infty}\ket{l}\bra{l}=1.
\end{equation}
We call the angular-momentum eigenbasis the $l$-basis.
Its Fourier transform is the $2\pi$-periodic $\varphi$-basis, 
\begin{equation}\label{fourier}
\ket{\varphi}=\sum_{l=-\infty}^{\infty}\ket{l}\mathrm{e}^{-\mathrm{i}l\varphi}
=\ket{\varphi+2\pi}.
\end{equation}
The orthogonality and the completeness of the $\varphi$-basis follow from
Eqs.~(\ref{ang mom}) and (\ref{fourier}), namely 
\begin{equation}\label{phi}
\braket{\varphi}{\varphi'}=2\pi\delta^{(2\pi)}\left(\varphi-\varphi'\right)
\quad\text{and } 
\int\limits_{(2\pi)}\!\frac{\mathrm{d}\varphi}{2\pi}
\ket{\varphi}\bra{\varphi}=1,
\end{equation}
where $\delta^{(2\pi)}(\ )$ is the $2\pi$-periodic delta function and the
integration covers any $2\pi$-interval. 
By construction, the $l$-basis and the $\varphi$-basis are unbiased: 
${|\braket{\varphi}{l}|^2=1}$ does not depend on the quantum numbers
$\varphi$ and $l$.

We can now introduce the unitary shift operator $E$ on the $l$-basis,
\begin{equation}
E\ket{l}=\ket{l+1}.
\end{equation}
Since the $l$-basis and the $\varphi$-basis are conjugate, the latter is
the eigenbasis of $E$, 
\begin{equation}
E\ket{\varphi}=\ket{\varphi}\ex{\varphi}.
\end{equation}
The shift operator $E$ and the angular-momentum operator $L$ are therefore
algebraically complete~\cite{englert06,schwinger01}; their algebraic
properties follow from the commutation relation
\begin{equation}\label{EL commutator}
  [L,E]=E.
\end{equation}
As mentioned in the Introduction, the Weyl--Heisenberg pair $(E,L)$ 
of the rotor can be obtained from a suitable ${d\to\infty}$ limit; 
for a textbook discussion, see Refs.~\cite{schwinger01,englert06}. 
 
We mentioned in the Introduction that, despite the similarities with
the linear motion, there is a fundamental difference: 
It is impossible to construct a third basis that is unbiased to both the
$l$-basis and the $\varphi$-basis. 
The nonexistence of a third basis can be seen as follows. 
Assume that there is a ket $\ket{x}$ belonging to such a basis, then the
property of being mutually unbiased requires that there are positive constants
$\lambda$ and $\mu$ such that 
\begin{equation}
\text{$\left|\braket{\varphi}{x}\right|^2=\lambda$
for all $\varphi$, and
$\left|\braket{l}{x}\right|^2=\mu$
for all $l$.}
\end{equation}
It then follows from the completeness relation in Eq.~(\ref{phi}) that
\begin{equation}
\braket{x}{x}=\bra{x}{\left(\,
\int\limits_{(2\pi)}\!\frac{\mathrm{d}\varphi}{2\pi}
\ket{\varphi}\bra{\varphi}\right)}
\ket{x}=\int\limits_{(2\pi)}\!\frac{\mathrm{d}\varphi}{2\pi}\lambda=\lambda.
\end{equation}
The other completeness relation in Eq.~(\ref{ang mom}), however, implies
\begin{equation}
\braket{x}{x}=\bra{x}\left(\sum_{l=-\infty}^{\infty}\ket{l}\bra{l}\right)\ket{x}
=\sum_{l=-\infty}^{\infty}\mu=\infty.
\end{equation}
The discrete spectrum of $L$ makes the series diverge and thus leads to a
contradiction. 

Therefore it remains unclear whether it is possible at all to obtain more than
two MUB for the rotor. In addition, we may wonder whether there is a continuous set of MUB as it
naturally obtains for all the other continuous degrees of freedom and whether
it is related to an underlying Heisenberg pair of complementary observables.

Since any Hilbert space whose dimension is countably infinite is isomorphic to the Hilbert space of motion along the line, for which a continuous set of MUB is known (see \cite{wootters87}, for example), there must be continuous sets of MUB for the rotor. Despite this mathematical insight,
the challenge is two-fold. Geometrically, we must find a mapping between the line and the rotor which respect the periodicity of the circular motion. And physically, this mapping should allow the expression of the Heisenberg pair $(Q,P)$ describing motion along the line and the Weyl--Heisenberg pair $(E,L)$ of the rotor in terms of each other.

We will examine two mappings. The first mapping is a stereographic mapping, which is not fully satisfactory: Geometrically, it provides a continuous set of MUB for the rotor, however, physically, there is no underlying Weyl--Heisenberg pair $(E,L)$. The second mapping exploits the one-to-one correspondence between natural numbers and integers, or in physical terms, between the Fock basis and the angular momentum basis. This mapping satisfies all the geometrical and physical requirements.

\section{A first continuous set of MUB}\label{sec 3}
Let us now consider the first mapping between the line and the rotor together with the corresponding set of MUB. It will turn out that the $\varphi$-basis is contained in this first continuous set of MUB for the rotor, whereas it is not contained in the second set of Sec.~\ref{sec 5} below. 

The continuous degree of freedom of linear motion admits a continuous set of
MUB. 
Geometrically, these MUB correspond to rotations of the position basis by an
angle $\theta$, which therefore labels the bases. 
Their wave functions take the simple form of a quadratic complex Gaussian
function 
\begin{eqnarray}\label{wf}
\Phi_y^{(\theta)}(q)=\frac{1}{\sqrt{\pi(1-\mathrm{e}^{2\mathrm{i}\theta})}} 
\exp\!{\left(\!\mathrm{i}\frac{qy}{\sin\theta}
-\frac{\mathrm{i}}{2}\frac{q^2+y^2}{\tan\theta}\right)} ,
\end{eqnarray}
where ${0 \le \theta < \pi}$ and the real parameter $y$ labels the basis
element \cite{note:BW}. 
First of all, for a given $\theta$, two wave functions $\Phi_y^{(\theta)}(q)$
and $\Phi_{y'}^{(\theta)}(q)$ are orthogonal, 
\begin{eqnarray}
\int\limits_{-\infty}^{\infty}\!\mathrm{d}q \, 
\Phi_y^{(\theta)}(q)^* \Phi_{y'}^{(\theta)}(q)= \delta(y-y'),
\end{eqnarray}
and we have the completeness relation
\begin{eqnarray}
\Psi(q)=\int\limits_{-\infty}^{\infty}\!\mathrm{d}y\, 
\Phi_y^{(\theta)}(q) \!\int\limits_{-\infty}^{\infty}\!\mathrm{d}q'\, 
\Phi_y^{(\theta)}(q')^* \Psi(q'),
\end{eqnarray}
for all wave functions $\Psi(q)$. 
Indeed, for a given $\theta$, the wave functions $\Phi_y^{(\theta)}(q)$ form a
basis. 
Second, any two bases $\theta_1$ and $\theta_2$, with $\theta_1 \neq
\theta_2,$ are unbiased: 
The modulus of the inner product between any wave function in the $\theta_1$
basis and any wave function in the $\theta_2$ basis is independent of the two
basis elements $y_1$ and $y_2$, 
\begin{equation}\label{MUBfirst}
\left|\int\limits_{-\infty}^{\infty}\!\mathrm{d}q\, 
\Phi_{y_1}^{(\theta_1)}(q)^* \Phi_{y_2}^{(\theta_2)}(q)\right|^2
=\frac{1}{2\pi\,|\sin(\theta_1-\theta_2)|}.
\end{equation}

Now, a simple change of variable readily provides a continuous set of MUB for
the rotor as specified by their wave functions in $\varphi$. 
For, the substitution $q=\tan (\varphi/2)$ allows us to write
\begin{eqnarray}
\int\limits_{-\infty}^{\infty}\!\!\mathrm{d}q \, \Phi_{y_1}^{(\theta_1)}\!(q)^* \, 
\Phi_{y_2}^{(\theta_2)}\!(q) 
=\! \int\limits_{(2\pi)}\!\frac{\mathrm{d}\varphi}{2\pi} \, 
\Gamma_{y_1}^{(\theta_1)}\!(\varphi)^* \, \Gamma_{y_2}^{(\theta_2)}\!(\varphi)
\end{eqnarray}
upon defining the $2\pi$-periodic wave functions
\begin{eqnarray}\label{def}
\Gamma_y^{(\theta)}(\varphi) =\sqrt{\frac{2\pi}{1+\cos\varphi}} \,
\Phi_y^{(\theta)}\bigl(\tan(\varphi/2)\bigr).
\end{eqnarray}
By construction, we conserve the three important properties of orthogonality,
completeness, and unbiasedness. 
It follows that the wave functions $\Gamma_y^{(\theta)}(\varphi)$ form a
continuous set of MUB for the rotor degree of freedom.

\begin{figure}
\centerline{\includegraphics[scale=1]{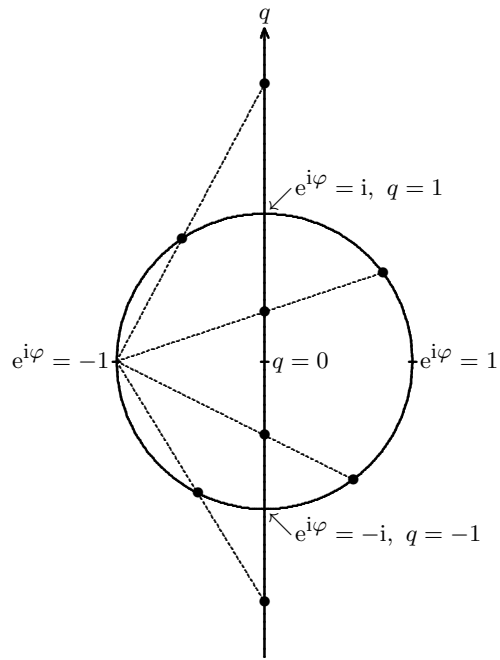}}
\caption{\label{qf-fig}%
Graphical representation of the change of variable $q=\tan(\varphi/2)$. 
This substitution is an example of stereographic projection: 
The unit circle is projected from the point 
${\mathrm{e}^{\mathrm{i}\varphi}=-1}$ 
onto the real line which intersects the circle at the two points 
$\mathrm{e}^{\mathrm{i} \varphi}= \pm \mathrm{i}$.  
The origin $q=0$ of the real line corresponds to the point 
$\mathrm{e}^{\mathrm{i} \varphi}=1$ on the circle. 
The dots $\bullet$ represent points on the circle and their stereographic
projection onto the real line. 
The dashed lines illustrate the imaginary line joining the origin of the
projection, the point on the circle to be projected and its projection onto
the real line.} 
\end{figure}

Consistent with the change of variable $q=\tan(\varphi/2)$, illustrated in
Fig.~\ref{qf-fig}, we would like to express the Weyl--Heisenberg pair $(E,L)$
of the circular motion and the Heisenberg pair $(Q,P)$ of the linear motion in
terms of each others. 
As noted in Ref.~\cite{debz10}, such a relation with the linear motion exists
for the two other continuous degrees of freedom of radial motion and motion
within a segment. 
However, it turns out that the present change of variable $q=\tan(\varphi/2)$
does not allow such a construction. 
The origin of the problem is the conflict between the substitution $q=\tan
(\varphi/2)$ and the $2\pi$-periodicity of the rotor degree of freedom. 
In particular, the limits ${q\to\infty}$ and ${q\to-\infty}$ both correspond
to ${\mathrm{e}^{\mathrm{i}\varphi}\to -1}$ although the ranges ${q \gg 1}$
and ${-q \gg 1}$ are not adjacent on the $q$ line. 
This eventually leads to an ill-defined Weyl--Heisenberg pair $(E,L)$
expressed in terms of the Heisenberg pair $(Q,P)$, while the inverse relation
does not present any issue; see the Appendix for more details. 

Although we obtained the present set (\ref{def}) of MUB in a rather
straightforward manner, we seek for another continuous set of MUB which would
not suffer from the lack of an underlying Heisenberg pair of complementary
observables. 
The primary reason is the following: Not only do we want to find a continuous
set of MUB for the rotor but we also want to settle the question whether the
rotor degree of freedom is on equal footing with the three other continuous
degrees of freedom. 
To do so, we must find an alternative set of MUB which arises from a bona fide
Heisenberg pair of complementary observables. 
This goal will be achieved by starting the construction from the angular
momentum instead of the angular position. 

We note for completeness that the basis for $\theta=0$ is essentially the
$\varphi$-basis of Eqs.~(\ref{fourier}) and (\ref{phi}), inasmuch as 
\begin{equation}\label{zero}
\int\limits_{(2\pi)}\!\frac{\mathrm{d}\varphi}{2\pi}\, 
\ket{\varphi} \Gamma_{y}^{(0)}(\varphi)
=\frac{\ket{\varphi=2 \arctan(y)}}{\sqrt{\pi (1+y^2)}}.
\end{equation}
Furthermore, when $\theta \neq 0$, the wave functions
$\Gamma^{(\theta)}_y(\varphi)$ of Eq.~(\ref{def}) have a pole at $\varphi=\pi$
and rapidly oscillate in the vicinity of that pole. 
The wave functions of the second continuous set of MUB of Sec.~\ref{sec 5}
below have similar singularities where, however, the angular position of the
pole will depend on the basis $\theta$.

\section{A Heisenberg pair for the rotor}\label{sec 4}
The construction of continuous MUB for the other continuous degrees of
freedom, given in Ref.~\cite{debz10},
relies on the respective Heisenberg pairs of complementary hermitian
observables, the analogs of position and momentum for motion along a line.
The procedure could be applied to the rotor degree of freedom as well 
if we had a Heisenberg pair for it, but that has been lacking, and the
construction of Sec.~\ref{sec 3} does not provide it. 

Owing to the discreteness of $l$ and the periodicity of $\varphi$, there is no
Heisenberg pair $(Q,P)$ for the rotor such that, say, $L$ is an invertible
function of $Q$ and $E$ is an invertible function of $P$.
We need to construct the Heisenberg pair in a different way.
One strategy is as follows.

\begin{figure}
\centerline{\includegraphics{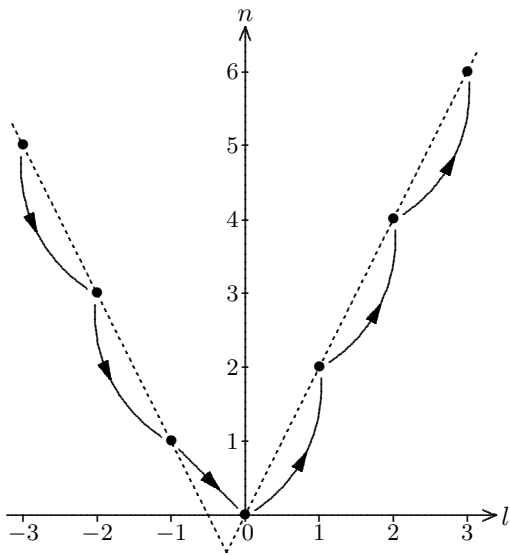}}
\caption{\label{nl-fig}%
Graphical representation of Eqs.~(\ref{Fock2})--(\ref{Fock4}).
The dashed line shows the relation of Eq.~(\ref{Fock2}) 
between the quantum numbers $l$ and $n$, with the dots $\bullet$ indicating
the integer pairs $(l,n)$ of physical significance. 
Negative $l$ values are mapped one-to-one onto odd $n$ values, whereas
nonnegative $l$ values are mapped onto even $n$ values.
The arrowed lines that connect them symbolize the mapping
${\ket{l}\to\ket{l+1}}$ associated with the unitary shift operator $E$ of
Eq.~(\ref{Fock4}).}   
\end{figure}

For position $Q$ and momentum $P$, we have the familiar Fock basis of kets
$\ket{n}$ with ${n=0,1,2,\dots}$, the eigenkets of the number operator 
${N=\frac{1}{2}(Q^2+P^2-1)}$,
\begin{equation}
  \label{Fock1}
  N\ket{n}=\ket{n}n.
\end{equation}
We identify the Fock basis with the $l$ basis in accordance with
\begin{equation}
  \label{Fock2}
  \ket{n}=\ket{l}\quad\text{if}\quad 2n+1=|4l+1|,
\end{equation}
which is illustrated in Fig.~\ref{nl-fig}.
It follows that $L$ and $N$ are functions of each other,
\begin{eqnarray}\label{Fock3}
  L&=&\frac{2N+1}{4}(-1)^N-\frac{1}{4},\nonumber\\
  N&=&\frac{1}{2}\bigl|4L+1\bigr|-\frac{1}{2}.
\end{eqnarray}
The unitary shift operator
\begin{eqnarray}
  \label{Fock4}
  E&=&\sum_{l=-\infty}^{\infty}\ket{l+1}\bra{l}\nonumber\\
&=&\sum_{\text{$n$ even}}\ket{n+2}\bra{n}
  +\sum_{\text{$n$ odd}}\ket{n}\bra{n+2}\nonumber\\
&&  +\ket{n=0}\bra{n=1}
\end{eqnarray}
can then be expressed with the aid of the isometric ladder operator for the
Fock states,
\begin{equation}
  \label{Fock5}
  A=\frac{1}{\sqrt{2N+2}}(Q+\mathrm{i}P)=\sum_{n=0}^{\infty}\ket{n}\bra{n+1},
\end{equation}
and its adjoint, for which $AA^{\dagger}=1$.
We have
\begin{equation}
  \label{Fock6}
  E={A^{\dagger}}^2\frac{1+(-1)^N}{2}+\frac{1-(-1)^N}{2}A^2+A-A^{\dagger}A^2,
\end{equation}
where ${A-A^{\dagger}A^2}={\ket{n=0}\bra{n=1}}={\ket{0}\bra{0}A}$ since
the projector on the sector with ${n=l=0}$ is the commutator of $A$ 
and $A^{\dagger}$
\begin{equation}
  \label{Fock7}
  [A,A^{\dagger}]=1-A^{\dagger}A=\ket{0}\bra{0}.
\end{equation}
In summary, in Eqs.~(\ref{Fock3}) and (\ref{Fock6}) we have the basic rotor
observables $E$ and $L$ expressed in terms of $N$, $A$, and $A^{\dagger}$
which are functions of the Heisenberg pair $(Q,P)$.

The reciprocal relations that state $Q$ and $P$ as functions of $E$ and $L$
are compactly written as
\begin{equation}
  \label{Fock8}
  Q+\mathrm{i}P=\sqrt{4L+2}\,\Pi_+RE+\sqrt{-4L}\,\Pi_-R.
\end{equation}
Here, $\Pi_+$ projects on the nonnegative $l$ values, and $\Pi_-$ on the
negative $l$ values,
\begin{equation}\label{Fock9}
\Pi_+=\sum_{l=0}^{\infty}\ket{l}\bra{l}, \qquad
\Pi_-=\sum_{l=-\infty}^{-1}\!\!\ket{l}\bra{l},
\end{equation}
and $R$ is
the hermitian and unitary reflection operator
\begin{equation}\label{Fock10}
R=\sum_{l=-\infty}^{\infty}\ket{l}\bra{-l}
=\sum_{l=-\infty}^{\infty}\ket{l}\bra{l}E^{2l}
=\sum_{l=-\infty}^{\infty}E^{-2l}\ket{l}\bra{l},
\end{equation}
which is such that ${Rf(E,L)=f(E^{\dagger},-L)R}$
holds for any operator function $f(E,L)$.
If one wishes, one can use
\begin{equation}
  \label{Fock11}
  \ket{l}\bra{l}
  =\int\limits_{(2\pi)}\!\frac{\mathrm{d}\alpha}{2\pi}\,\ex{(L-l)\alpha},
\end{equation}
or other identities of this kind, to state more explicit functions 
of $L$ for $\Pi_{\pm}$ and $R$.
It is a matter of inspection to verify that $[Q,P]=\mathrm{i}$ for the
hermitian $(Q,P)$ pair defined by Eq.~(\ref{Fock8}).

The fundamental difference between the construction here and that in
Sec.~\ref{sec 3} (with details in the Appendix) should be obvious: 
In Sec.~\ref{sec 3}, we are employing the one-to-one mapping of
Fig.~\ref{qf-fig} between the circle with one point removed and the real line,
whereas we are now relying on the one-to-one mapping of Fig.~\ref{nl-fig}
between integers and natural numbers.

\section{A second continuous set of MUB}\label{sec 5}
With the Heisenberg pair of Eq.~(\ref{Fock8}) at hand, we follow the usual
procedure and note that any two linear combinations $\alpha Q+\beta P$ and
$\alpha' Q+\beta' P$ are a pair of complementary observables if
${\alpha\beta'\neq\alpha'\beta}$ holds for the real coefficients; 
see, for instance, Sec.~1.1.8 in Ref.~\cite{debz10}.
We restrict ourselves to the one-parameter set with $\alpha=\cos\theta$ and
$\beta=\sin\theta$ for ${0\leq\theta<\pi}$,
\begin{equation}
  \label{MUB1}
  Y_\theta\equiv Q\cos\theta+P\sin\theta
=\mathrm{e}^{\mathrm{i}\theta N}Q\,\mathrm{e}^{-\mathrm{i}\theta N}.
\end{equation}
The eigenkets $\ket{\theta;y}$
of $Y_\theta$ are then given in terms of the eigenkets $\ket{q}$ of $Q$,
\begin{equation}
  \label{MUB2}
  Y_\theta\ket{\theta;y}=\ket{\theta;y}y\quad\text{for}\quad
\ket{\theta;y}=\mathrm{e}^{\mathrm{i}\theta N}\ket{q=y}.
\end{equation}
For each $\theta$, the $\ket{\theta;y}$s make up a continuous basis of kets,
and the bases for different $\theta$ values are unbiased: For
${\theta_1\neq\theta_2}$, the transition probability density 
\begin{equation}
  \label{MUB3}
  \bigl|\braket{\theta_1;y_1}{\theta_2;y_2}\bigr|^2
=\frac{1}{2\pi\,|\sin(\theta_1-\theta_2)|}
\end{equation}
does not depend on the quantum numbers $y_1$ and $y_2$.

In passing we note that the similarity between Eqs.~(\ref{MUB3}) and
(\ref{MUBfirst}) is, of course, not accidental. 
In fact, we have $\Phi_y^{(\theta)}(q)=\braket{q}{\theta;y}$ but the
geometrical meaning of the $q$-basis here is quite different from that of the
$q$-basis in Sec.~\ref{sec 3}.

\begin{figure}
\centerline{\includegraphics{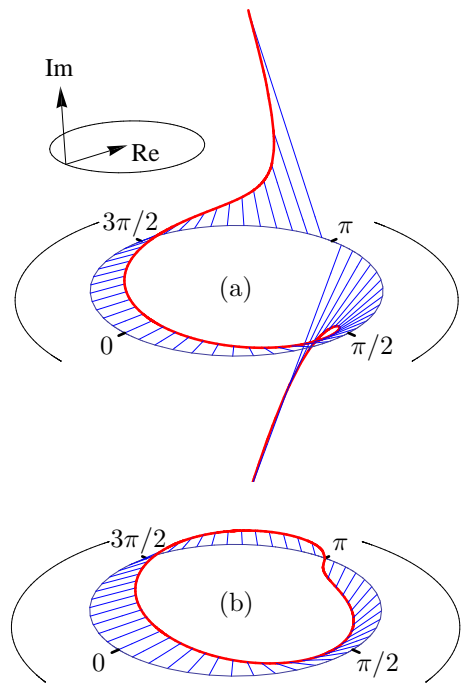}}
\caption{\label{psi-fig1}%
(color online) 
The wave functions $\psi^{(0)}_y(\varphi)$ for ${y=0}$. 
One $2\pi$-period of $\varphi$ is represented by a circle. 
At each point on the circle, we have a complex plane perpendicular to the
plane of the circle, with the real axis toward the center of the $\varphi$
circle. 
In these complex planes we mark the values of the wave functions by thin blue
lines, whose end points make up the thick red lines. 
The unit distance in the complex planes is indicated by the outside arcs for
${\pi/2<\varphi<\pi}$ and ${3\pi/2<\varphi<2\pi}$, which mark points with
${\psi=-1}$. 
Plot (a) shows $\psi^{(0)}_0(\varphi)$ which has a simple pole at
${\varphi=\pi}$. 
After removing the pole $1/\sqrt{1+\cos\varphi}$, plot (b) shows the smooth
function $\chi^{(+)}_{0}(\varphi)$ of Eq.~(\ref{MUB8}). 
}
\end{figure}

The well-known position wave functions for the Fock states,
\begin{equation}
  \label{MUB4}
\braket{q}{n}=
\pi^{-\frac{1}{4}}\left(2^nn!\right)^{-\frac{1}{2}}\mathrm{e}^{-\frac{1}{2}q^2}H_n(q)
\equiv f_n(q),  
\end{equation}
where $H_n(q)$ denotes the $n$th Hermite polynomial, translate into the
wave function of $\ket{\theta;y}$ in the $l$-basis,
\begin{equation}
  \label{MUB5}
  \braket{l}{\theta;y}=\ex{n\theta}f_n(y)
\Bigr|_{\mbox{\footnotesize$n=\frac{1}{2}|4l+1|-\frac{1}{2}$}}.
\end{equation}
The periodic wave function in the $\varphi$-basis is then
available in terms of the Fourier sum
\begin{eqnarray}
  \label{MUB6}
\psi^{(\theta)}_y(\varphi)\equiv
  \braket{\varphi}{\theta;y}
&=& \sum_{l=0}^{\infty}\Bigl[\ex{l(\varphi+2\theta)}f_{2l}(y)\\\nonumber
&&\hphantom{\sum_{l=0}^{\infty}\Bigl[}
+\mathrm{e}^{-\mathrm{i}\theta}
\mathrm{e}^{-\mathrm{i}(l+1)(\varphi-2\theta)}f_{2l+1}(y)\Bigr]
\end{eqnarray}
that is implied by Eqs.~(\ref{ang mom}) and (\ref{fourier}).
Since the identity
\begin{eqnarray}\label{MUB7}
  \psi^{(\theta)}_y(\varphi)&=&\frac{1}{2}\biggl(
\psi^{(0)}_y(\varphi+2\theta)+\psi^{(0)}_{-y}(\varphi+2\theta)\\\nonumber
&&\hphantom{\frac{1}{2}\biggl(}
+\mathrm{e}^{-\mathrm{i}\theta}\Bigl[\psi^{(0)}_y(\varphi-2\theta)
-\psi^{(0)}_{-y}(\varphi-2\theta)\Bigr]\biggr)
\end{eqnarray}
expresses the wave functions of the $\theta$-basis in terms of those for 
${\theta=0}$, one needs to evaluate the series in Eq.~(\ref{MUB6}) 
only for ${\theta=0}$.

\begin{figure*}
\centerline{\includegraphics{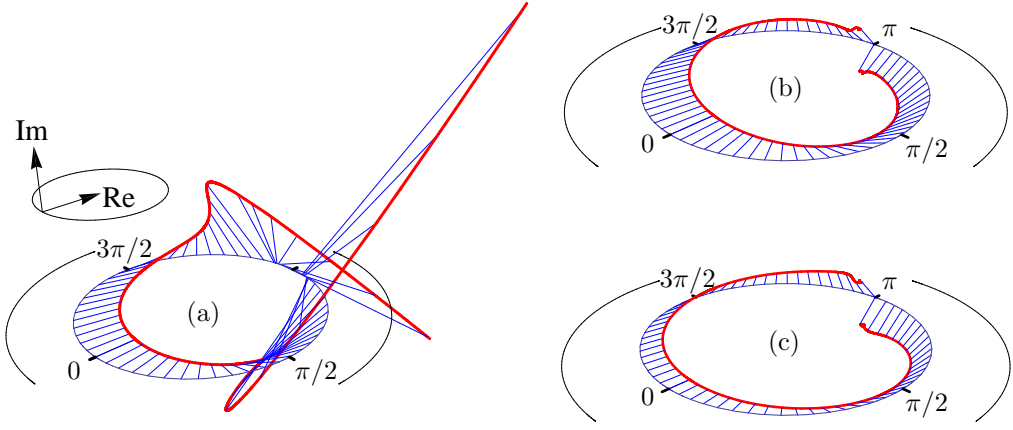}}
\caption{\label{psi-fig2}%
(color online) 
Plot (a) shows the wave function $\psi^{(0)}_{\frac{1}{2}}(\varphi)$. 
The vicinity of ${\varphi=\pi}$ is excluded because this wave function is
oscillating very rapidly there. 
After removing the pole $1/\sqrt{1+\cos\varphi}$ and the rapidly oscillating
factors $\mathrm{e}^{\pm\frac{\mathrm{i}}{8}\tan\frac{\varphi}{2}}$, we have
the even-in-$y$ and odd-in-$y$ parts $\chi^{(\pm)}_{\frac{1}{2}}(\varphi)$ of
Eqs.~(\ref{MUB8}) and (\ref{MUB9}), which are shown in plots (b) and (c). 
These functions have remaining low-amplitude oscillations in the vicinity of
$\varphi=\pi$ but no poles at ${\varphi=\pi}$. 
However, the imaginary part of $\chi^{(\pm)}_{\frac{1}{2}}(\varphi)$ is
discontinuous at $\varphi=\pi$. 
}
\end{figure*}

For illustration, Figs.~\ref{psi-fig1}(a) and \ref{psi-fig2}(a)
show $\psi^{(0)}_y(\varphi)$ for ${y=0}$ and ${y=1/2}$.
These wave functions are singular at ${\varphi=\pi}$: $\psi^{(0)}_0(\varphi)$
has a pole there, whereas $\psi^{(0)}_{\frac{1}{2}}(\varphi)$ is finite but
oscillates arbitrarily rapidly in the vicinity of ${\varphi=\pi}$, which is a
common feature of all wave functions $\psi^{(0)}_y(\varphi)$ with ${y\neq0}$. 
The pole and the rapidly oscillating factors are exhibited in the even-in-$y$
and odd-in-$y$ parts of $\psi^{(0)}_y(\varphi)$,
\begin{eqnarray}\label{MUB8} 
\frac{1}{2}\Bigl[\psi^{(0)}_y(\varphi)+\psi^{(0)}_{-y}(\varphi)\Bigr] 
&=&\sum_{l=0}^{\infty}\ex{l\varphi}f_{2l}(y)
\nonumber\\
&=&\frac{\mathrm{e}^{\frac{\mathrm{i}}{2}y^2\tan\frac{\varphi}{2}}}
        {\sqrt{1+\cos \varphi}}\chi^{(+)}_y(\varphi)
\end{eqnarray}
and
\begin{eqnarray}\label{MUB9}
\frac{1}{2}\Bigl[\psi^{(0)}_y(\varphi)-\psi^{(0)}_{-y}(\varphi)\Bigr]
&=&\sum_{l=0}^{\infty}\mathrm{e}^{-\mathrm{i}(l+1)\varphi}f_{2l+1}(y)
\nonumber\\
&=&\frac{\mathrm{e}^{-\frac{\mathrm{i}}{2}y^2\tan\frac{\varphi}{2}}}
        {\sqrt{1+\cos \varphi}}\chi^{(-)}_y(\varphi)\,,
\end{eqnarray}
where the factors $\chi^{(\pm)}_y(\varphi)$ are smooth functions of $\varphi$
with remaining small-amplitude oscillations around $\varphi=\pi$ but no poles
at $\varphi=\pi$. 
For ${y=0}$, we have ${\chi^{(-)}_0(\varphi)=0}$. 
Figure~\ref{psi-fig1}(b) is a plot of $\chi^{(+)}_0(\varphi)$ while 
Figs.~\ref{psi-fig2}(b) and (c) are plots of $\chi^{(\pm)}_{\frac{1}{2}}(\varphi)$.

\section{Summary}
We provided two continuous sets of MUB for the rotor degree of freedom. 
We thus answered the question of whether there are more than two MUB for the
rotor degree of freedom by providing explicit continuous sets. 
These two sets of MUB are found by mapping the problem of finding MUB for the
rotor onto that of the linear motion, for which a method of constructing a
continuous set of MUB is known. 
The first continuous set is specified by simple wave functions but is not
satisfactory as it does not relate to an underlying Heisenberg pair. 
So, we established such a Heisenberg pair of complementary observables for the
rotor to construct a second and more suitable continuous set of MUB. 
In summary, the rotor degree of freedom is on equal footing with the other
continuous degrees of freedom: 
For all of them there are continuous sets of MUB which are related to an
underlying Heisenberg pair of complementary observables. 

The Heisenberg pair of Eq.~(\ref{Fock8}) is a mathematical construct that
serves our purpose well but, admittedly, we are not aware of another rotor
problem in which these operators would appear naturally and thus reveal their
physical significance. 
Conversely, we do not know whether the unitary operator of Eq.~(\ref{Fock6}), 
regarded as an observable for a linear degree of motion, such as a harmonic
oscillator, is relevant in another context.

\acknowledgments
We thank Jun Suzuki for valuable discussions.
This work is supported by the National Research Foundation and the Ministry of
Education, Singapore.

\appendix

\section*{Appendix}

In this appendix we demonstrate why the construction of Sec.~\ref{sec 3} is
not fully satisfactory. 

Following the change of variable $q=\tan(\varphi/2)$, we would like to express
the Heisenberg pair $(Q,P)$ and the Weyl--Heisenberg pair $(E,L)$ in terms of
each other. 
Of course, to be valid, these operators must have all the right properties of
hermiticity and self-adjointness as well as the right spectrum. 

First, let us find the expressions of the two hermitian operators $Q$ and $P$
in terms of $E$ and $L$. 
According to Eq.~(\ref{def}), we express the $2\pi$-periodic eigenbras
$\bra{\varphi}$ of $E$ in terms of the eigenbras $\bra{q}$ of $Q$ as 
\begin{equation}\label{varphi}
 \bra{\varphi}=\sqrt{\frac{2\pi}{1+\cos\varphi}}\,\bra{q=\tan(\varphi/2)}.
\end{equation}
The position operator $Q$ is given by $\bra{q}Q=q\bra{q}$, or equivalently,  
$\bra{\varphi}Q=\tan(\varphi/2) \bra{\varphi}$, so that
\begin{eqnarray}\label{qqdef}
Q=\mathrm{i}\frac{1-E}{1+E}.
\end{eqnarray}
Next we want to find its conjugate operator $P$. 
For this purpose, we consider the unitary shift operator 
$\mathrm{e}^{\mathrm{i} a P}$ with real $a$, such that
\begin{eqnarray}
\bra{q}  \mathrm{e}^{\mathrm{i} a P}  = \bra{q+a}.
\end{eqnarray}
To obtain its expression in terms of $E$ and $L$, we look at its action on a
bra $\bra{\varphi}$. 
It reads
\begin{eqnarray}
\bra{\varphi}\mathrm{e}^{\mathrm{i} a P} 
= \sqrt{\frac{\mathrm{d}\varphi'}{\mathrm{d}\varphi}} \bra{\varphi'},
\end{eqnarray}
where
\begin{eqnarray}
\varphi'=2 \arctan\bigl(\tan(\varphi/2)+a\bigr).
\end{eqnarray}
The resulting $E;L$-ordered form of the shift operator is
\begin{eqnarray}
\mathrm{e}^{\mathrm{i} a P}=\frac{1}{|1-\mathrm{i}\frac{a}{2}(1+E)|}
{\left(\frac{1+\mathrm{i}\frac{a}{2}(1+E^\dagger)}
{1-\mathrm{i}\frac{a}{2}(1+E)}\right)}^L.
\end{eqnarray}
We also have $\mathrm{e}^{\mathrm{i} 0 P}=1$ as well as the group property 
${\mathrm{e}^{\mathrm{i} a P} \mathrm{e}^{\mathrm{i} b P}
=\mathrm{e}^{\mathrm{i} (a+b) P}}$. 
We now consider the $a\to0$ limit of $\mathrm{e}^{\mathrm{i} a P}$ to obtain
its generator $P$ and find 
\begin{eqnarray}\label{qqqdef}
P=\frac{1}{2}\bigl|1+E\bigr| \,L\,\bigl|1+E\bigr|,
\end{eqnarray}
where ${|A|=\sqrt{A^\dagger A}}$ for any operator $A$. 
As required, $P$ is hermitian and we verify that the commutation relation
between $Q$ and $P$ indeed is $[Q,P]=\mathrm{i}$. 
It remains to look at the spectral properties of $P$ to conclude that we have
constructed a well-defined Heisenberg pair of complementary observables
$(Q,P)$. 
The $\varphi$ wave functions of the eigenkets of $P$ are given by
\begin{eqnarray}
\sqrt{1+\cos \varphi} \, \braket{\varphi}{p}
= c\, \mathrm{e}^{\mathrm{i} p \tan(\varphi/2)},
\end{eqnarray}
where $c$ is a normalization constant. 
The choice ${c=1}$ together with the definition~(\ref{varphi}) imply the
expected Fourier coefficient 
\begin{eqnarray} \label{fourcoef}
\braket{q}{p}=\frac{1}{\sqrt{2\pi}} \mathrm{e}^{\mathrm{i} p q}.
\end{eqnarray}
Therefore the two operators $Q$ and $P$, expressed in terms of the
Weyl--Heisenberg pair $(E,L)$, represent a valid Heisenberg pair of
complementary observables: 
They have the right Heisenberg commutation relation as well as the right
properties. 
Let us now focus on the two operators $E$ and $L$ in terms of $Q$ and $P$.

First of all, we invert Eqs.~(\ref{qqdef}) and (\ref{qqqdef}) to obtain
\begin{equation}
E=\frac{1+\mathrm{i}Q}{1-\mathrm{i}Q} \label{E}
\end{equation}
and
\begin{equation}
L=\frac{1}{2}\sqrt{1+Q^2}P\sqrt{1+Q^2} \label{L}.
\end{equation}
These two hermitian operators yield the right commutation relation $[L,E]=E$. 
As earlier, we must also check that the operators have the right spectrum. 
By construction, the eigenvalues of $E$ are the phase factors 
${\mathrm{e}^{\mathrm{i}\varphi}=(1+\mathrm{i}q)/(1-\mathrm{i}q)}$
and, upon inverting Eq.~(\ref{varphi}) 
[cf.\ Eq.~(\ref{zero})] 
\begin{equation}\label{qdef}
\bra{q}=\frac{\bra{\varphi=2\arctan(q)}}{\sqrt{\pi(1+q^2)}},
\end{equation}
we confirm Eq.~(\ref{E}).
Let us now investigate the spectral properties of the seemingly unproblematic
hermitian operator $L$ that is defined by the $(Q,P)$ function in
Eq.~(\ref{L}). 
The $q$ wave functions of its eigenkets are
\begin{eqnarray}
\braket{q}{\lambda} = \frac{c'}{\sqrt{1+q^2}} 
\Bigl(\frac{1+ \mathrm{i} q}{1- \mathrm{i} q}\Bigr)^\lambda,
\end{eqnarray}
where the eigenvalue $\lambda$ is \emph{any} real number, not restricted to
integers, and $c'$ is a normalization constant. 
That all real numbers are eigenvalues is also evident as soon as one realizes
that the unitary transformation ${Q\to Q}$, ${P \to P+2x/(1+Q^2)}$ adds $x$ to
the right-hand side of Eq.~(\ref{L}), whereby $x$ can be any real number. 
It follows that the $L$ operator of Eq.~(\ref{L}) is not the $L$ operator of
Sec.~\ref{sec 2}, the generator of the unitary cyclic shift 
${\bra{\varphi}\to\bra{\varphi+\alpha}}$. 

In conjunction with Eq.~(\ref{qdef}), the choice $c'=1/\sqrt{\pi}$ gives
the $\varphi$ wave functions 
\begin{eqnarray}\label{fourcoef2}
\braket{\varphi}{\lambda}
=\mathrm{e}^{\mathrm{i} \lambda \bigl( \varphi - 2\pi\lfloor\frac{\varphi}{2\pi}\rceil\bigr)},
\end{eqnarray}
where  $\lfloor x \rceil$ denotes the integer that is nearest to $x$. 
Furthermore, the eigenvectors of the $L$ of Eq.~(\ref{L}) are not all
orthogonal. 
Indeed, we have
\begin{eqnarray}
\braket{\lambda}{\lambda'} 
= \mathrm{sinc}\bigl( \pi (\lambda-\lambda') \bigr),
\end{eqnarray}
so that only the eigenvectors whose eigenvalues differ by an integer are
orthogonal. 
Consequently, the $\lambda$-basis is overcomplete: 
There are many completeness relations, such as
\begin{eqnarray}\label{complet}
\sum_{l=-\infty}^{\infty}\ket{l+\lambda_0}\bra{l+\lambda_0 } =1,
\end{eqnarray}
with ${0 \leq \lambda_0 < 1}$, say. 
Mathematically speaking, the operator $L$ of Eq.~(\ref{L}) is hermitian but
not self-adjoint. 

We may wonder whether the above issues remain if we start from the unitary
shift operator $\mathrm{e}^{\mathrm{i} \alpha L}$ instead of inverting
Eq.~(\ref{qqqdef}). 
We proceed from the expression of the $2\pi$-periodic $\varphi$ bras
$\bra{\varphi}$ in terms of the $q$ bras $\bra{q}$ in Eq.~(\ref{varphi}).
The unitary shift $\mathrm{e}^{\mathrm{i} \alpha L}$ acts on $\bra{\varphi}$ as
\begin{eqnarray}
\bra{\varphi}\mathrm{e}^{\mathrm{i} \alpha L}  = \bra{\varphi+\alpha}.
\end{eqnarray}
On a bra $\bra{q}$, it then reads
\begin{eqnarray}\label{qp1}
\bra{q}\mathrm{e}^{\mathrm{i} \alpha L}  
= \sqrt{\frac{\mathrm{d}q'}{\mathrm{d}q}} \bra{q'},
\end{eqnarray}
where
\begin{eqnarray}\label{qp2}
q'=\frac{q \cos(\alpha/2)+\sin(\alpha/2)}{\cos(\alpha/2)-q \sin(\alpha/2)}.
\end{eqnarray}
From Eqs.~(\ref{qp1}) and (\ref{qp2}),  we derive the $Q;P$-ordered form of
the shift operator $\mathrm{e}^{\mathrm{i}\alpha L}$, which is 
\begin{eqnarray}\label{shift}
 \mathrm{e}^{\mathrm{i}\alpha L}&=&\frac{1}{|\cos(\alpha/2)-Q\sin(\alpha/2)|}
\\\nonumber
&&\mbox{}\times \exp{\left(\mathrm{i}\frac{(1+Q^2)\sin(\alpha/2)}
                                         {\cos(\alpha/2)-Q\sin(\alpha/2)} 
                    P\right)}.
\end{eqnarray}
It follows that $\mathrm{e}^{\mathrm{i} 0 L}=1$ and 
$\mathrm{e}^{\mathrm{i} \alpha L} \mathrm{e}^{\mathrm{i} \beta L}
=\mathrm{e}^{\mathrm{i} (\alpha+\beta) L}$. 
Therefore the $Q;P$-ordered form of the unitary shift 
$\mathrm{e}^{\mathrm{i} \alpha L}$ is well-defined. 
Moreover, $\mathrm{e}^{\mathrm{i} 2\pi L}=1$ tells us that the eigenvalues of
$L$ are integers. 
However, while this unitary shift is well-defined, it does not admit a uniform
$\alpha \rightarrow 0$ limit and, therefore, it does not have a self-adjoint
generator. 

The problem is that the substitution $q=\tan (\varphi/2)$ breaks the
periodicity between the two end points with $\pm\pi$. 
The consequences can be seen at various occasions, for example when we look at
the spectrum of the operator $L$ of Eq.~(\ref{L}) or notice the lack of a
generator for the unitary shift $\mathrm{e}^{\mathrm{i} \alpha L}$ of
Eq.~(\ref{qp1}).

As a puzzling remark, and quite independent of the rotor degree of freedom,
let us point out that the hermitian operator $L$ of Eq.~(\ref{L}) can be
written as the commutator 
\begin{equation}
L= - \mathrm{i} \biggl[\frac{Q}{2}+\frac{Q^3}{6},\frac{P^2}{2}\biggr],
\end{equation}
where the right-hand side is the time derivative of the observable 
$Z=Q/2+Q^3/6$ under the evolution governed by a Hamiltonian of the familiar
form ${H=P^2/2+V(Q)}$ with some potential energy $V(Q)$. 
We thus observe that, although the observable $Z$ is surely self-adjoint, its
time derivative is not!


\begin{thebibliography}{99}
\bibitem{note:pairs}
Strictly speaking, a Weyl pair is a pair of unitary observables with a
Weyl commutation rule, and if one can associate hermitian generators with
the unitary operators, then one has a Heisenberg pair of hermitian
operators. 
In the case of the rotor, we have the mixed $(E,L)$ pair of Sec.~\ref{sec 2}, 
with unitary $E$ and hermitian $L$.
\bibitem{note:conventions}
Throughout the paper, operators are denoted by letters in upper case,
and lower-case letters denote numbers.
\bibitem{note:BW}
Up to a $q$-independent phase factor, these wave functions are the ones used by Wootters \cite{wootters87}.
\end{thebibliography}
\end{document}